\documentstyle[fullpage,epic,eepic,11pt,fancyheadings]{article}

\input{psfig}

\newtheorem{teo}{Theorem}[section]
\newtheorem{lemma}{Lemma}[section]
\newtheorem{defin}[teo]{Definition}

\newcommand{\dissquorum}[1]{\mbox{$\lceil (#1 + t + 1)/2 \rceil$}}
\newcommand{\onethird}[1]{\mbox{$\lfloor (#1-1)/3 \rfloor$}}

\newcommand{\remove}[1]{}
\newcommand{\act}[1]{\mbox{{\sl active}$_{#1}$}}

\newcommand{\echo}{${\sl E}~$}

\newcommand{\witness}[1]{{\sl witness}\mbox{$(#1)$}}

\newcommand{\sign}[2]{\mbox{$#2_{K_#1}$}}
\newcommand{\target}[1]{{\sl peers}{\mbox{$_{#1}$}}}
\newcommand{\wt}{\mbox{$W_{3T}$}}
\newcommand{\wact}{\mbox{$W_{active}$}}

\newcommand{\la}{\mbox{\tt <}}
\newcommand{\ra}{\mbox{\tt >}}

\newenvironment{proof}{\noindent{\bf Proof :\/}}{$\Box$\vskip 0.1in}

\begin{document}
\setlength{\baselineskip}{15pt}
\bibliographystyle{plain}

\title{\Large\bf Secure Reliable Multicast Protocols in a WAN\thanks{
Preprint of a paper to appear in the Distributed Computing Journal.}}

\author{ 
\begin{tabular}[t]{c@{\extracolsep{0.2in}}c@{\extracolsep{0.2in}}c}
Dahlia Malkhi & Michael Merritt & Ohad Rodeh\\ \\ 
\multicolumn{2}{c}{AT\&T Shannon Labs, New Jersey} &
	The Hebrew University of Jerusalem\\
\multicolumn{2}{c}{\tt \{dalia,mischu\}@research.att.com} &
	{\tt orodeh@cs.huji.ac.il}
\end{tabular}
}
\date{}

\maketitle

\rhead{}
\lhead{}
\setlength{\headrulewidth}{0pt}
\cfoot{\footnotesize\sl Revision 1, Submitted to the Journal of
Distributed Computing.}
\thispagestyle{fancy}

\subsection*{\centering Abstract}
{\em A secure reliable multicast protocol enables a process to send a
message to a group of recipients such that all correct destinations
receive the same message, despite the malicious efforts of fewer than
a third of the total number of processes, including the sender. This has been shown to be a
useful tool in building secure distributed services, albeit with a
cost that typically grows linearly with the size of the system.  For
very large networks, for which this is prohibitive, we
present two approaches for reducing the cost: First, we show a
protocol whose cost is on the order of the number of tolerated
failures. Secondly, we show how relaxing the consistency requirement
to a probabilistic guarantee can reduce the
associated cost, effectively to a constant.  }

\section{Introduction}
\label{intro}

Communication over a large and sparse internet is a challenging
problem because communication links experience diverse delays and
failures.  Moreover, in a wide area network (WAN), security is most
crucial, since communicating parties are geographically dispersed and
thus are more prone to attacks.  The problem addressed in this paper
is the secure reliable multicast problem, namely, how to distribute
messages among a large group of participants so that all the
(correctly behaving) participants agree on messages' contents, despite
the malicious cooperation of up to a third of the members.

Experience with building robust distributed systems proves that
(secure) reliable multicast is an important tool for distributed
applications. Distributed platforms can increase the efficiency of
services and diminish the trust put in each component.  For example,
the Omega key management system~\cite{RFLW96} provides key backup,
recovery and other functions in a penetration-tolerant way using the
Rampart distributed communication infrastructure~\cite{Rei95}.  
In such a service and others, distribution might
increase the sensitivity to failures and malicious attacks. To address
issues of availability and security, distributed services must rely on
mechanisms for maintaining consistent intermediate state and for
making coordinated decisions. Reliable multicast underlies the
mechanisms used in many infrastructure tools supporting such
distributed systems (for a representative collection,
cf.~\cite{cacm-gcs}).

Previous work on the reliable multicast problem suffers from message
complexity and computation costs that do not scale to very large
communication groups: Toueg's echo\_broadcast~\cite{Tou84,BT85}
requires $O(n^2)$ authenticated message exchanges for each message
delivery (where $n$ is the size of the group). Reiter improved this
message complexity in the ECHO protocol of the Rampart
system~\cite{Rei94} through the usage of digital signatures. The ECHO
protocol incurs $O(n)$ {\em signed\/} message exchanges, and thus,
message complexity is improved at the expense of increased computation
cost.  Malkhi and Reiter~\cite{MR96} extended this approach to
amortize the cost of computing digital signatures over multiple
messages through a technique called {\em acknowledgment chaining\/},
where a signed acknowledgment directly verifies the message it
acknowledges and {\em indirectly}, every message that message
acknowledges.  Nevertheless, $O(n)$ digital signatures are in the
critical path between message sending and its delivery. For a very
large group of hundreds or thousands of members, this may be
prohibitive.  

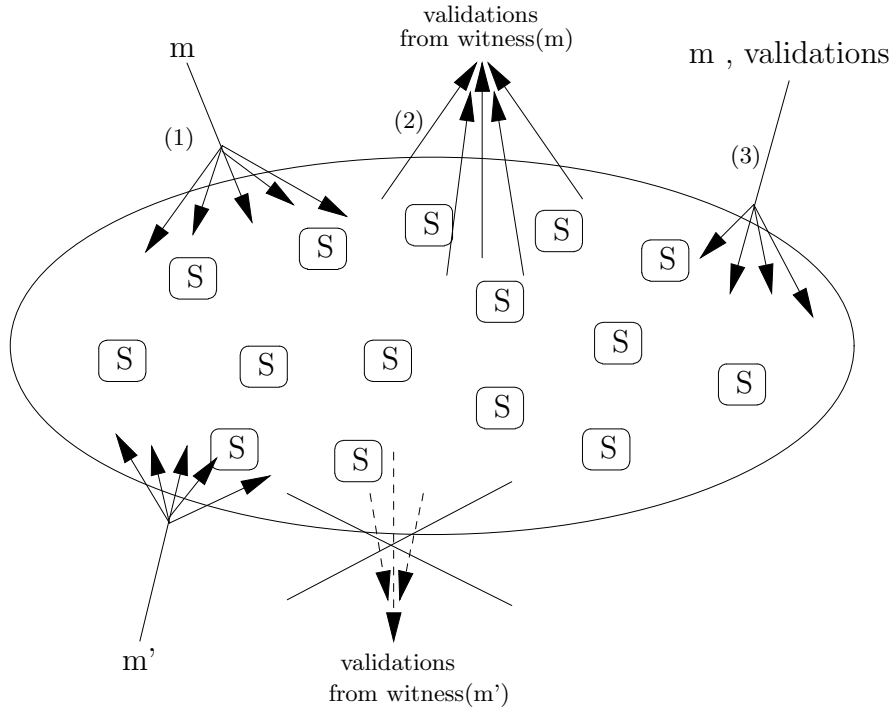
\begin{figure}[htb]
\centerline{\setlength{\unitlength}{0.00041667in}
\begingroup\makeatletter\ifx\SetFigFont\undefined%
\gdef\SetFigFont#1#2#3#4#5{%
  \reset@font\fontsize{#1}{#2pt}%
  \fontfamily{#3}\fontseries{#4}\fontshape{#5}%
  \selectfont}%
\fi\endgroup%
{\renewcommand{\dashlinestretch}{30}
\begin{picture}(10855,8862)(0,-10)
\put(1245,4221){\arc{210}{1.5708}{3.1416}}
\put(1245,4536){\arc{210}{3.1416}{4.7124}}
\put(1635,4536){\arc{210}{4.7124}{6.2832}}
\put(1635,4221){\arc{210}{0}{1.5708}}
\path(1140,4221)(1140,4536)
\path(1245,4641)(1635,4641)
\path(1740,4536)(1740,4221)
\path(1635,4116)(1245,4116)
\put(1365,4266){\makebox(0,0)[lb]{\smash{{{\SetFigFont{12}{14.4}{\rmdefault}{\mddefault}{\updefault}S}}}}}
\put(2145,5271){\arc{210}{1.5708}{3.1416}}
\put(2145,5586){\arc{210}{3.1416}{4.7124}}
\put(2535,5586){\arc{210}{4.7124}{6.2832}}
\put(2535,5271){\arc{210}{0}{1.5708}}
\path(2040,5271)(2040,5586)
\path(2145,5691)(2535,5691)
\path(2640,5586)(2640,5271)
\path(2535,5166)(2145,5166)
\put(2265,5316){\makebox(0,0)[lb]{\smash{{{\SetFigFont{12}{14.4}{\rmdefault}{\mddefault}{\updefault}S}}}}}
\put(3045,4146){\arc{210}{1.5708}{3.1416}}
\put(3045,4461){\arc{210}{3.1416}{4.7124}}
\put(3435,4461){\arc{210}{4.7124}{6.2832}}
\put(3435,4146){\arc{210}{0}{1.5708}}
\path(2940,4146)(2940,4461)
\path(3045,4566)(3435,4566)
\path(3540,4461)(3540,4146)
\path(3435,4041)(3045,4041)
\put(3165,4191){\makebox(0,0)[lb]{\smash{{{\SetFigFont{12}{14.4}{\rmdefault}{\mddefault}{\updefault}S}}}}}
\put(2670,3096){\arc{210}{1.5708}{3.1416}}
\put(2670,3411){\arc{210}{3.1416}{4.7124}}
\put(3060,3411){\arc{210}{4.7124}{6.2832}}
\put(3060,3096){\arc{210}{0}{1.5708}}
\path(2565,3096)(2565,3411)
\path(2670,3516)(3060,3516)
\path(3165,3411)(3165,3096)
\path(3060,2991)(2670,2991)
\put(2790,3141){\makebox(0,0)[lb]{\smash{{{\SetFigFont{12}{14.4}{\rmdefault}{\mddefault}{\updefault}S}}}}}
\put(4245,2946){\arc{210}{1.5708}{3.1416}}
\put(4245,3261){\arc{210}{3.1416}{4.7124}}
\put(4635,3261){\arc{210}{4.7124}{6.2832}}
\put(4635,2946){\arc{210}{0}{1.5708}}
\path(4140,2946)(4140,3261)
\path(4245,3366)(4635,3366)
\path(4740,3261)(4740,2946)
\path(4635,2841)(4245,2841)
\put(4365,2991){\makebox(0,0)[lb]{\smash{{{\SetFigFont{12}{14.4}{\rmdefault}{\mddefault}{\updefault}S}}}}}
\put(4620,4221){\arc{210}{1.5708}{3.1416}}
\put(4620,4536){\arc{210}{3.1416}{4.7124}}
\put(5010,4536){\arc{210}{4.7124}{6.2832}}
\put(5010,4221){\arc{210}{0}{1.5708}}
\path(4515,4221)(4515,4536)
\path(4620,4641)(5010,4641)
\path(5115,4536)(5115,4221)
\path(5010,4116)(4620,4116)
\put(4740,4266){\makebox(0,0)[lb]{\smash{{{\SetFigFont{12}{14.4}{\rmdefault}{\mddefault}{\updefault}S}}}}}
\put(3795,5646){\arc{210}{1.5708}{3.1416}}
\put(3795,5961){\arc{210}{3.1416}{4.7124}}
\put(4185,5961){\arc{210}{4.7124}{6.2832}}
\put(4185,5646){\arc{210}{0}{1.5708}}
\path(3690,5646)(3690,5961)
\path(3795,6066)(4185,6066)
\path(4290,5961)(4290,5646)
\path(4185,5541)(3795,5541)
\put(3915,5691){\makebox(0,0)[lb]{\smash{{{\SetFigFont{12}{14.4}{\rmdefault}{\mddefault}{\updefault}S}}}}}
\put(5145,5946){\arc{210}{1.5708}{3.1416}}
\put(5145,6261){\arc{210}{3.1416}{4.7124}}
\put(5535,6261){\arc{210}{4.7124}{6.2832}}
\put(5535,5946){\arc{210}{0}{1.5708}}
\path(5040,5946)(5040,6261)
\path(5145,6366)(5535,6366)
\path(5640,6261)(5640,5946)
\path(5535,5841)(5145,5841)
\put(5265,5991){\makebox(0,0)[lb]{\smash{{{\SetFigFont{12}{14.4}{\rmdefault}{\mddefault}{\updefault}S}}}}}
\put(6045,4971){\arc{210}{1.5708}{3.1416}}
\put(6045,5286){\arc{210}{3.1416}{4.7124}}
\put(6435,5286){\arc{210}{4.7124}{6.2832}}
\put(6435,4971){\arc{210}{0}{1.5708}}
\path(5940,4971)(5940,5286)
\path(6045,5391)(6435,5391)
\path(6540,5286)(6540,4971)
\path(6435,4866)(6045,4866)
\put(6165,5016){\makebox(0,0)[lb]{\smash{{{\SetFigFont{12}{14.4}{\rmdefault}{\mddefault}{\updefault}S}}}}}
\put(6045,3621){\arc{210}{1.5708}{3.1416}}
\put(6045,3936){\arc{210}{3.1416}{4.7124}}
\put(6435,3936){\arc{210}{4.7124}{6.2832}}
\put(6435,3621){\arc{210}{0}{1.5708}}
\path(5940,3621)(5940,3936)
\path(6045,4041)(6435,4041)
\path(6540,3936)(6540,3621)
\path(6435,3516)(6045,3516)
\put(6165,3666){\makebox(0,0)[lb]{\smash{{{\SetFigFont{12}{14.4}{\rmdefault}{\mddefault}{\updefault}S}}}}}
\put(7395,3096){\arc{210}{1.5708}{3.1416}}
\put(7395,3411){\arc{210}{3.1416}{4.7124}}
\put(7785,3411){\arc{210}{4.7124}{6.2832}}
\put(7785,3096){\arc{210}{0}{1.5708}}
\path(7290,3096)(7290,3411)
\path(7395,3516)(7785,3516)
\path(7890,3411)(7890,3096)
\path(7785,2991)(7395,2991)
\put(7515,3141){\makebox(0,0)[lb]{\smash{{{\SetFigFont{12}{14.4}{\rmdefault}{\mddefault}{\updefault}S}}}}}
\put(9120,3921){\arc{210}{1.5708}{3.1416}}
\put(9120,4236){\arc{210}{3.1416}{4.7124}}
\put(9510,4236){\arc{210}{4.7124}{6.2832}}
\put(9510,3921){\arc{210}{0}{1.5708}}
\path(9015,3921)(9015,4236)
\path(9120,4341)(9510,4341)
\path(9615,4236)(9615,3921)
\path(9510,3816)(9120,3816)
\put(9240,3966){\makebox(0,0)[lb]{\smash{{{\SetFigFont{12}{14.4}{\rmdefault}{\mddefault}{\updefault}S}}}}}
\put(8145,5496){\arc{210}{1.5708}{3.1416}}
\put(8145,5811){\arc{210}{3.1416}{4.7124}}
\put(8535,5811){\arc{210}{4.7124}{6.2832}}
\put(8535,5496){\arc{210}{0}{1.5708}}
\path(8040,5496)(8040,5811)
\path(8145,5916)(8535,5916)
\path(8640,5811)(8640,5496)
\path(8535,5391)(8145,5391)
\put(8265,5541){\makebox(0,0)[lb]{\smash{{{\SetFigFont{12}{14.4}{\rmdefault}{\mddefault}{\updefault}S}}}}}
\put(7545,4446){\arc{210}{1.5708}{3.1416}}
\put(7545,4761){\arc{210}{3.1416}{4.7124}}
\put(7935,4761){\arc{210}{4.7124}{6.2832}}
\put(7935,4446){\arc{210}{0}{1.5708}}
\path(7440,4446)(7440,4761)
\path(7545,4866)(7935,4866)
\path(8040,4761)(8040,4446)
\path(7935,4341)(7545,4341)
\put(7665,4491){\makebox(0,0)[lb]{\smash{{{\SetFigFont{12}{14.4}{\rmdefault}{\mddefault}{\updefault}S}}}}}
\put(6795,5871){\arc{210}{1.5708}{3.1416}}
\put(6795,6186){\arc{210}{3.1416}{4.7124}}
\put(7185,6186){\arc{210}{4.7124}{6.2832}}
\put(7185,5871){\arc{210}{0}{1.5708}}
\path(6690,5871)(6690,6186)
\path(6795,6291)(7185,6291)
\path(7290,6186)(7290,5871)
\path(7185,5766)(6795,5766)
\put(6915,5916){\makebox(0,0)[lb]{\smash{{{\SetFigFont{12}{14.4}{\rmdefault}{\mddefault}{\updefault}S}}}}}
\put(5377,4566){\ellipse{10724}{4800}}
\path(2265,8166)(3090,6141)
\blacken\path(2870.825,6440.437)(3090.000,6141.000)(3037.521,6508.350)(2870.825,6440.437)
\path(6015,5691)(6015,8166)
\blacken\path(6105.000,7806.000)(6015.000,8166.000)(5925.000,7806.000)(6105.000,7806.000)
\path(1665,816)(2265,3291)
\blacken\path(2267.651,2919.930)(2265.000,3291.000)(2092.717,2962.338)(2267.651,2919.930)
\dashline{120.000}(4890,3216)(4890,816)
\blacken\path(4800.000,1176.000)(4890.000,816.000)(4980.000,1176.000)(4800.000,1176.000)
\path(3540,2691)(6390,1266)
\path(6390,2841)(3540,1341)
\path(2715,7116)(2340,5991)
\blacken\path(2368.460,6360.986)(2340.000,5991.000)(2539.223,6304.065)(2368.460,6360.986)
\path(2715,7041)(3615,6366)
\blacken\path(3273.000,6510.000)(3615.000,6366.000)(3381.000,6654.000)(3273.000,6510.000)
\path(2715,7116)(4290,6216)
\blacken\path(3932.780,6316.468)(4290.000,6216.000)(4022.085,6472.752)(3932.780,6316.468)
\path(2715,7116)(1740,5766)
\blacken\path(1877.815,6110.539)(1740.000,5766.000)(2023.738,6005.150)(1877.815,6110.539)
\path(9915,7941)(9165,5241)
\blacken\path(9174.635,5611.954)(9165.000,5241.000)(9348.068,5563.778)(9174.635,5611.954)
\path(9465,6366)(9690,5241)
\blacken\path(9531.146,5576.359)(9690.000,5241.000)(9707.650,5611.659)(9531.146,5576.359)
\path(9465,6366)(8790,5691)
\blacken\path(8980.919,6009.198)(8790.000,5691.000)(9108.198,5881.919)(8980.919,6009.198)
\path(9465,6366)(10215,4941)
\blacken\path(9967.689,5217.653)(10215.000,4941.000)(10126.974,5301.488)(9967.689,5217.653)
\path(7290,6441)(6090,8166)
\blacken\path(6369.465,7921.870)(6090.000,8166.000)(6221.702,7819.078)(6369.465,7921.870)
\path(6540,5466)(6165,7791)
\blacken\path(6311.175,7449.924)(6165.000,7791.000)(6133.472,7421.262)(6311.175,7449.924)
\path(5565,5466)(5865,7791)
\blacken\path(5908.190,7422.443)(5865.000,7791.000)(5729.670,7445.477)(5908.190,7422.443)
\path(4740,6441)(5940,8166)
\blacken\path(5808.298,7819.078)(5940.000,8166.000)(5660.535,7921.870)(5808.298,7819.078)
\path(2040,2316)(1815,3291)
\blacken\path(1983.645,2960.457)(1815.000,3291.000)(1808.254,2919.982)(1983.645,2960.457)
\path(2040,2391)(2640,3141)
\blacken\path(2485.388,2803.665)(2640.000,3141.000)(2344.832,2916.110)(2485.388,2803.665)
\path(2040,2316)(1365,3441)
\blacken\path(1627.393,3178.607)(1365.000,3441.000)(1473.044,3085.998)(1627.393,3178.607)
\path(2040,2316)(3315,2916)
\blacken\path(3027.587,2681.279)(3315.000,2916.000)(2950.944,2844.147)(3027.587,2681.279)
\dashline{120.000}(5265,2691)(4965,1341)
\blacken\path(4955.238,1711.951)(4965.000,1341.000)(5130.952,1672.904)(4955.238,1711.951)
\dashline{120.000}(4590,2691)(4815,1341)
\blacken\path(4667.041,1681.306)(4815.000,1341.000)(4844.592,1710.898)(4667.041,1681.306)
\put(2040,8241){\makebox(0,0)[lb]{\smash{{{\SetFigFont{12}{14.4}{\rmdefault}{\mddefault}{\updefault}m}}}}}
\put(1440,441){\makebox(0,0)[lb]{\smash{{{\SetFigFont{12}{14.4}{\rmdefault}{\mddefault}{\updefault}m'}}}}}
\put(4215,441){\makebox(0,0)[lb]{\smash{{{\SetFigFont{9}{10.8}{\rmdefault}{\mddefault}{\updefault}validations}}}}}
\put(4065,51){\makebox(0,0)[lb]{\smash{{{\SetFigFont{9}{10.8}{\rmdefault}{\mddefault}{\updefault}from witness(m')}}}}}
\put(8640,8166){\makebox(0,0)[lb]{\smash{{{\SetFigFont{12}{14.4}{\rmdefault}{\mddefault}{\updefault}m , validations}}}}}
\put(5265,8691){\makebox(0,0)[lb]{\smash{{{\SetFigFont{9}{10.8}{\rmdefault}{\mddefault}{\updefault}validations}}}}}
\put(4965,8376){\makebox(0,0)[lb]{\smash{{{\SetFigFont{9}{10.8}{\rmdefault}{\mddefault}{\updefault}from witness(m)}}}}}
\put(1965,7116){\makebox(0,0)[lb]{\smash{{{\SetFigFont{9}{10.8}{\rmdefault}{\mddefault}{\updefault}(1)}}}}}
\put(4890,7341){\makebox(0,0)[lb]{\smash{{{\SetFigFont{9}{10.8}{\rmdefault}{\mddefault}{\updefault}(2)}}}}}
\put(9165,6891){\makebox(0,0)[lb]{\smash{{{\SetFigFont{9}{10.8}{\rmdefault}{\mddefault}{\updefault}(3)}}}}}
\end{picture}
}}
\caption{Framework of Secure Reliable Multicast Protocols}\label{fig:E}
\end{figure}

In this paper, we propose two approaches for reducing the cost and
delay associated with reliable multicast. First, we show a protocol
whose cost is on the order of the number of tolerated failures, rather
than of the group size. Secondly, we show how relaxing the consistency
requirement to a selected probabilistic guarantee can reduce the
associated cost to a constant.  The principle underlying agreement on
message contents in previous works, as well as ours, is the following
(see Figure~\ref{fig:E}): For a process $p$ to send a message $m$ to
the group of processes, signed validations are obtained for $m$
from a certain set of processes, thereby enabling delivery of $m$.  We
call this validation set the {\em witness set\/} of $m$, denoted
$witness(m)$.  Witness sets are chosen so that any pair of them
intersect at a correct process, and such that some witness set is
always accessible despite failures. More precisely, witness sets
satisfy the {\em Consistency\/} and {\em Availability\/} requirements
of {\em Byzantine dissemination quorum systems} (cf.~\cite{MR98}), as
follows:

\begin{defin}
A dissemination quorum system is a set of subsets, called quorums,
satisfying: \\
For every set $B$ of faulty processes, and every two
quorums $Q_1, Q_2$, $Q_1 \cap Q_2 \not\subseteq B$ ({\bf
Consistency}). \\
For every set $B$ of faulty processes, there
exists a quorum $Q$ such that $Q \subseteq \overline{B}$ ({\bf
Availability}).
\end{defin}

Using dissemination quorums as witness sets for messages, if a faulty
process generates two messages $m$, $m'$ with the same sender and
sequence number and different contents (called {\em conflicting
messages}), the corresponding witness sets intersect at a correct
process (by Consistency). Thus, they cannot both obtain validations,
and at most one of them will be delivered by the correct processes of
the system.  Further, dissemination quorums ensure availability, such
that a correct process can always obtain validation from a witness set
despite possible failures. For a resilience threshold $t <
\onethird{n}$, previous works used quorums of size
\dissquorum{n}, providing both consistency and availability.

    Our first improvement in the {\em 3T\/} protocol drops the quorum
size from $\dissquorum{n}$ to $2t+1$.  When $t$ is a small constant,
this improvement is substantial, since we need only wait for $O(t)$
processes, no matter how big the WAN might be.  Briefly, the trick in
bringing down the size of the witness sets is in designating for every
message $m$ a witness set $\wt(m)$ of size $3t+1$, determined by its
sender and sequence number. A message must get validations from $2t+1$
processes, out of the designated set of $3t+1$, in order to be
delivered.

    Our second improvement stems from relaxing the requirement on
processes to (always) agree on messages' contents to a {\em
probabilistic\/} requirement. This leads to a protocol that consists
of two-regimes: The first one, called the {\em no-failure\/} regime,
is applied in faultless scenarios. It is very efficient, incurring
only a {\bf constant} overhead in message exchanges and signature
computing.  The second regime is the {\em recovery\/} regime, and is
resorted to in case of failures.  The two regimes inter-operate by
having the witnesses of the no-failure regime actively probe the
system to detect conflicting messages.  We introduce a probabilistic
protocol combining both regimes, \act{t}, whose properties are as
follows:

\begin{itemize}
	\item Given a resilience threshold $t$, \act{t} can be tuned
	to guarantee agreement on messages contents by all the correct
	processes on all but an arbitrarily small expected fraction $\epsilon$
	of the messages. Those messages that might be subject to
	conflicting delivery are determined by the random choices made
	by the processes after the execution starts, and hence a
	non-adaptive adversary cannot effect their choice.

	\item The overhead of forming agreement on message contents
	in \act{t} in faultless circumstances is determined by two
	constants that depend on $\epsilon$ only (and not on the
	system size or $t$).

\end{itemize}

In this paper, we assume a static set of communicating processes. It
is possible, however, to use known techniques (e.g., in the group
communication context one can use \cite{Rei94}) to
extend our protocols to operate in a dynamic environment in which
processes may leave or join the set of destination processes and in
which processes may fail and recover.

The rest of this paper is organized as follows: In
Section~\ref{model} we formally present our assumptions about the
system. Section~\ref{problem} presents a precise problem definition,
and demonstrates feasibility through a simple solution.
Section~\ref{3T} contains the {\em 3T} protocol description, and
Section~\ref{active} contains the \act{t} protocol description.  In
Section~\ref{load} we analyze the {\em load\/} induced on
processes participating in our protocols. We
conclude in Section~\ref{conclusion}.

\section{Model}
\label{model}

The system contains $n$ participating processes, denoted $P = \{ p_1, p_2,
\ldots, p_n \}$, up to $t \leq \onethird{n}$ of which may be
arbitrarily (Byzantine) faulty.  A faulty process may deviate from the
behavior dictated by the protocol in an arbitrary way, subject to
cryptographic assumptions.  
(Such failures are called authenticated {\em Byzantine\/} failures.)

Processes interact solely through message passing.  Every pair of
correct processes is connected via an authenticated FIFO channel, that
guarantees the identity of senders using any one of well known
cryptographic techniques.  We assume no limit on the relative speeds
of different processes or a known upper bound on message transmission
delays. However, we assume that every message sent between two
processes has a known probability of reaching its destination, which
grows to one as the elapsed time from sending increases. The last
property is needed only in the \act{t} protocol, and ensures the
delays can be set in the protocol to guarantee that fault notification
reaches all correct processes. In practice, this can be realized using
quality guaranteed out-of-band communication for control messages.

A system that admits Byzantine failures is often abstracted as having
an {\em adversary\/} working against the successful execution of
protocols.  We assume the following limitations on the adversary's
powers: The adversary chooses which processes are faulty at the
beginning of the execution, and thus its choice is non-adaptive.
Every process possesses a private key, known only to itself,
that may be used for signing data using a known public key
cryptographic method (such as~\cite{RSA78}).  Let $d$ be any data
block. We denote by \sign{i}{d} the signature of $p_i$ on the data $d$
by means of $p_i$'s private key.  We assume that every process in the
system may obtain the public keys of all of the other processes, such
that it can verify the authenticity of signatures.  The adversary
cannot access the local memories of the correct processes or the
communication among them, nor break their private keys, and thus
cannot obtain data internal to computations in the protocols.

Our protocols also make use of a cryptographically secure hash
function $H$ (such as MD5~\cite{MD5}).  Our assumption is that it is
computationally infeasible for the adversary to find two different
messages $m$ and $m'$ such that $H(m)=H(m')$.

\section{The Problem Definition and a Basic Solution}
\label{problem}

\noindent
A reliable multicast protocol provides each process $p$ with two operations:

\begin{description}
\item[{\sl WAN-multicast(m)}:] $p$ sends the multicast message $m$ to the group.

\item[{\sl WAN-deliver(m)}:] $p$ delivers a multicast message $m$,
making it available to applications at $p$.
\end{description}

\noindent
For convenience, we assume that a multicast message $m$ contains several fields:

\begin{description}
\item{{\sl sender}$(m)$:} The identity of the
sending process.
\item{{\sl seq}$(m)$:} A count of the multicast messages originated by sender.
\item{{\sl payload}$(m)$:} The (opaque) data of the message.
\end{description}

The protocol should guarantee that all of the correct members of the group
agree on the delivered messages, and furthermore, that messages sent
by correct processes are (eventually) delivered by all of the correct
processes. More precisely, the protocol should maintain the following
properties:

\begin{description}

\item[Integrity:] Let $p$ be a correct process. 
Then for any sequence number $s$,
$p$ performs {\sl WAN-deliver}$(m)$ for a message $m$ with {\sl seq}$(m) = s$
at most once, and if {\sl sender}$(m)$ is correct, 
then only if {\sl sender}$(m)$ executed {\sl WAN-multicast}$(m)$.

\item[Self-delivery:] Let $p$ be a correct process. If $p$
executes {\sl WAN-multicast}$(m)$ then eventually $p$ delivers $m$, {\em
i.e.}, eventually $p$ executes {\sl WAN-deliver}$(m)$.

\item[Reliability:] Let $p_i$ and $p_j$ be two correct processes. If
$p_i$ delivers a message from $p_k$ with sequence number {\sl seq}
(via {\sl WAN-deliver}$(m)$), then (eventually) $p_j$ delivers a message
from $p_k$ with sequence number {\sl seq}.

\item[(Probabilistic) Agreement:] Let $p_i$ and $p_j$ be two correct
processes. Let $p_i$ deliver a message $m$, $p_j$ deliver $m'$, such
that ${\sl sender}(m) = {\sl sender}(m')$ and ${\sl seq}(m) = {\sl
seq}(m')$. Then (with very high probability) $p_i$ and $p_j$ delivered
the same message, {\em i.e.}, {\sl payload}$(m)$ = {\sl payload}$(m')$

\end{description}

The problem statement above is strictly weaker than the Byzantine
agreement problem~\cite{LSP82}, which is known to be unsolvable in
asynchronous systems~\cite{FLP}. This statement holds even if we use
the unconditional Agreement requirement.  The reason is that only
messages from correct processes are required to be delivered, and thus
messages from faulty processes can ``hang'' forever. Note that there
is no ordering requirement among different messages, and thus the
problem statement is weaker than the totally ordered reliable
multicast problem, which can be solved only
probabilistically~\cite{MMA90,MM95}. The reliable multicast problem is
solvable in our environment, as is demonstrated by the \echo protocol
depicted in Figure~\ref{echo} (which borrows from the Rampart ECHO
multicast protocol~\cite{Rei94}).  Throughout the protocol, each
process $p_i$ maintains a delivery vector $delivery_i[]$ containing
the sequence number of the last WAN-delivered message from every other
process. $delivery_i[]$ is initially set to zero.

This protocol assumes the presence of a {\em stability mechanism\/},
(SM), utilized by the processes, that allows each process to learn
when a message has been delivered by other processes, for purposes of
re-transmission and garbage collection.  The details of such a
mechanism are omitted (for the details of such a mechanism, in the
context of a group communication system, see
e.g. \cite{birman-lightwt}).  However, we note that by properly tuning
timeout periods and by packing multiple messages together (e.g., by
piggybacking on regular traffic), the cost of such a mechanism is
negligible in practice. The mechanism must assure the following
properties:

\begin{description}

\item[SM\_Reliability:] Let $p_i$ and $p_j$ be two correct processes.
If $p_i$ performs {\sl WAN-deliver}$(m)$,
then eventually $p_j$ knows that $p_i$ performed {\sl WAN-deliver}$(m)$.

\item[SM\_Integrity:] Let $p_i$ and $p_j$ be two correct processes. 
If $p_j$ learns from the stability mechanism that $p_i$ performed {\sl WAN-deliver}$(m)$,
then indeed, $p_i$ performed {\sl WAN-deliver}$(m)$.

\end{description}

Protocols can be used as components in more complex protocols;
to separate the messages of disparate protocols, each contains an initial field
indicating to which protocol it belongs.  Messages within each protocol
similarly contain fields indicating their role in the protocol.
(E.g. as acknowledgements.)

\begin{figure}[hbt]
\noindent
\rule{\columnwidth}{.5mm}

\begin{enumerate}
\item
For a process $p_i$ to {\sl WAN-multicast} message $m$, (such that
{\sl sender}$(m) = p_i$), and $p_i$ has previously sent messages up to
sequence number ${\sl seq}(m) - 1$, process $p_i$ sends
\[\la {\tt E}, {\tt regular}, p_i, {\sl seq}(m), H(m) \ra\] 
to every process in $P$, and waits to obtain $A = \{\sign{j}{\la {\tt E}, {\tt ack},
p_i, {\sl seq}(m), H(m) \ra} ~|~ p_j \in P' \}$, a set of signed
acknowledgments from any set $P'$ of \dissquorum{n} distinct
processes.  It then sends the following
to every process in $P$:
\[\la {\tt E}, {\tt deliver}, m, A \ra~.\] 

\item 
When $p_i$ receives a message $\la {\tt E}, {\tt regular}, p_j, cnt, h \ra$
from $p_j$, and no conflicting message was previously received from $p_j$, then $p_i$
sends back to $p_j$ a signed acknowledgment 
\[\sign{i}{\la {\tt E}, {\tt ack}, p_j, cnt, h \ra}~.\]

\item 
When $p_i$ receives a message $\la {\tt E}, {\tt deliver}, m, A \ra$,
such that $A$ contains a valid set of acknowledgments for $\la {\sl
sender}(m), {\sl seq}(m), H(m) \ra$, (acknowledgements for $m$ from
\dissquorum{n} distinct processes) and such that $delivery_i[{\sl
sender}(m)] = {\sl seq}(m) - 1$, $p_i$ performs {\sl WAN-deliver}$(m)$
and sets $delivery_i[{\sl sender}(m)]$ to ${\sl seq}(m)$.  If a
timeout period has passed and $p_j$ is not known to have delivered
$m$, $p_i$ sends $\la {\tt E}, {\tt deliver}, m, A \ra$ to $p_j$.
\end{enumerate}

\rule{\columnwidth}{.5mm}
\caption{The \echo protocol}
\label{echo}
\end{figure}

The \echo protocol ensures secure reliable multicast. However, it is
inefficient in faultless runs, incurring an overhead (in addition to
$O(n)$ transmissions for multicast) of $O(n)$ signatures and message
exchanges per delivery; this might be an intolerable overhead for very
large groups.  We shall improve it in the next section.

We now proceed to verify that the \echo protocol satisfies Integrity,
Self-delivery, Reliability and Agreement. This is the basic
proof in this paper; while simple and straightforward, it facilitates
later proofs of optimizations.

\begin{defin}
Two acknowledgements, 
$\sign{i}{\la {\tt E}, {\tt ack}, p_k, cnt_k, h_k \ra}$
and
$\sign{j}{\la {\tt E}, {\tt ack}, p_\ell, cnt_\ell, h_\ell \ra}$,
{\em conflict} if 
$p_k = p_\ell$ and 
$cnt_k = cnt_\ell$, but
$h_k \neq h_\ell$.
\end{defin}

In proving the security of the \echo protocol, we will make use of the
fact that in any run of the protocol,
no correct process multicasts conflicting messages and 
no correct process signs conflicting acknowledgements.
In addition, the lemma below states several properties that relate
acknowledgement sets to the corresponding message transmissions:

\begin{lemma}
\label{lemma1}
In any run of the \echo protocol, the following hold:
\begin{enumerate}
\item
\label{lemma_1}
If process $p$ is correct in a run, then
a correct process signs an acknowledgement 
for a message $m$ with {\sl sender}($m$) = $p$
only if $p$ multicasts message $m$.
\item
\label{lemma_2}
If process $p$ is correct in a run, and the run contains a set of valid
{\em E\/} acknowledgements for a message $m$ with {\sl sender}($m$) =
$p$, then $p$ {\sl WAN-multicast} message $m$ in the run.
\item
\label{lemma_3}
The run does not contain two valid sets of conflicting {\em E\/}
acknowledgements.
\end{enumerate}
\end{lemma}

\begin{proof}
A correct process $q$ signs an acknowledgement for $m$ by a correct
sender only when it
receives $m$ over an authenticated channel from {\sl sender}($m$). The
first property then follows from the fact that a signed
acknowledgement from $q$ of the form
$\sign{q}{\la {\tt E}, {\tt ack}, p, cnt, H(m) \ra}$ contains the identity of
the sender $p$ = {\sl sender}($m$).

To see that the second property holds, recall that a valid
acknowledgement set contains acknowledgements from \dissquorum{n}
distinct processes, which must contain at least $t+1$ processes.
Hence, a valid set of acknowledgements must contain an acknowledgement
signed by a correct process.  Since $p$ is correct, by property
(\ref{lemma_1}) of the lemma, $m$ was multicast by $p$.

Finally, note that two sets of valid acknowledgements must intersect
in at least one correct process. Since a correct process never signs
conflicting acknowledgements, the third property follows.
\end{proof}

\begin{teo}(Integrity)
Let $p_i$ be a correct process participating in the \echo
protocol. Then for any message $m$, $p_i$ performs {\sl WAN-deliver}$(m)$ 
at most once, and
if {\sl sender}$(m)$ is correct, then only if {\sl sender}$(m)$
executed {\sl WAN-multicast}$(m)$.
\label{echoint}
\end{teo}
\begin{proof}
That $p_i$ suppresses duplicate deliveries is immediate from the
protocol. It is left to show that if {\sl sender}$(m)$ is correct, it
must have sent $m$.  To prove this fact, consider that for $p_i$ to
deliver $m$, $p_i$ must have obtained a set $A$ of valid
acknowledgments for $m$. By Lemma~\ref{lemma1}(\ref{lemma_2}), it
follows that $m$ must have been sent by {\sl sender}$(m)$.
\end{proof}

\begin{teo}(Self-delivery)
Let $p_i$ be a correct process participating in the \echo protocol. If
$p_i$ executes {\sl WAN-multicast}(m) then eventually $p_i$ executes {\sl
WAN-deliver}(m).
\label{echoself}
\end{teo}

\begin{proof}
Notice that $\dissquorum{n} \le n-t$, thus there are at least
\dissquorum{n} correct processes in $P$. Thus, if $p_i$ sends $\la {\tt
E}, {\tt regular}, p_i, {\sl seq}(m), H(m) \ra$ to every process in
$P$, then at least \dissquorum{n} correct processes $p_j$ will receive
it.  Since no correct process receives a conflicting message from
$p_i$, each will acknowledge it, sending back a $\sign{j}{\la {\tt E},
{\tt ack}, p_i, {\sl seq}(m), H(m) \ra}$ message to $p_i$, thereby
enabling delivery of $m$ by $p_i$.
\end{proof}

\begin{teo}(Reliability)
Let $p_i$ and $p_j$ be two correct processes participating in the \echo
protocol. If $p_i$ performs {\sl WAN-deliver}($m$), 
then eventually $p_j$ performs {\sl WAN-deliver}($m$).
\label{echorel}
\end{teo}

\begin{proof}
For $p_i$ to deliver $m$, $p_i$ must have obtained a set $A$ of valid
acknowledgments for $m$ from \dissquorum{n} processes.  If $p_i$
learns that $p_j$ delivered $m$, then by SM\_Integrity we are done.
Alternatively, if $p_i$ does not learn that $p_j$ delivered $m$, then
after a timeout period $p_i$ sends $\la {\tt E}, {\tt deliver}, m, A
\ra$ to $p_j$.  By Lemma~\ref{lemma1}(\ref{lemma_3}), $p_j$ cannot
have received a conflicting set of acknowledgements, so at the latest,
upon receipt of $\la {\tt E}, {\tt deliver}, m, A \ra$ from $p_i$,
process $p_j$ performs {\sl WAN-deliver}($m$).
\end{proof}

\begin{teo}(Agreement)
Let $p_i$ and $p_j$ be two correct processes participating in the \echo
protocol. Let $p_i$ deliver a message $m$, $p_j$ deliver $m'$, such that
${\sl sender}(m) = {\sl sender}(m')$ and ${\sl seq}(m) = {\sl
seq}(m')$. Then $p_i$ and $p_j$ delivered the
same message, {\em i.e.}, ${\sl payload}(m) = {\sl payload}(m')$.
\label{echoagr}
\end{teo}

\begin{proof}
For $p_i$ to deliver $m$, $p_i$ must have obtained a set $A$
of valid acknowledgments for $m$ from \dissquorum{n} processes in
$P$. Likewise, $p_j$ must have obtained a set $A'$ of
\dissquorum{n} valid acknowledgments for $m'$. 
By Lemma~\ref{lemma1}(\ref{lemma_3}), $A$ and $A'$ do not conflict,
and by the security of $H$, $m=m'$.
\end{proof}

\section{The 3T Protocol}
\label{3T}

In this section we introduce the {\em 3T\/} protocol.  The improvement
in this protocol over the \echo protocol above comes from designating
a potential witness set for each message $m$ based on the pair $\la${\sl
sender}$(m),$ {\sl seq}$(m) \ra$.  The choice of potential witness set for
$p_i$'s $k$'th message is determined by a function $\wt(p_i, k)$,
whose range is the set of subsets of exactly $(3t+1)$ distinct process
id's.  For simplicity, we denote $\wt(m) = \wt({\sl sender}(m), {\sl
seq}(m))$.  For efficiency, $\wt$ could be chosen to distribute the
load of witnessing over distinct sets of processes for different
messages.  Figure~\ref{fig:3t} provides the details of the {\em 3T\/}
protocol.

\begin{figure}[tbh]
\rule{\columnwidth}{.5mm}
\begin{enumerate}

\item For a process $p_i$ to perform {\sl WAN-multicast}($m$), (such
that {\sl sender}$(m) = p_i$), and $p_i$ has previously sent messages
up to sequence number ${\sl seq}(m) - 1$, $p_i$ sends
\[ \la {\tt 3T}, {\tt regular}, p_i, {\sl seq}(m), H(m) \ra
\] 
to every process in $\wt(m)$, and waits to obtain 
$A = \{ \sign{j}{\la {\tt 3T}, {\tt ack}, 
p_i, {\sl seq}(m), H(m) \ra} ~|~ p_j \in P' \}$, 
a set of signed acknowledgments from any subset $P'$ of $\wt(m)$
comprising of $2t+1$ distinct processes.
Process $p_i$ then sends to every process in $P$
\[ \la {\tt 3T}, {\tt deliver}, m, A \ra~.
\]

\item When $p_i$ receives a message $\la {\tt 3T}, {\tt regular}, p_j,
cnt, h \ra$ from $p_j$, such that no conflicting message was
previously received, $p_i$ sends to $p_j$ a signed acknowledgment 
\[ \sign{i}{\la {\tt 3T}, {\tt ack}, p_j, cnt, h \ra}~.
\] 

\item When $p_i$ receives a message $\la {\tt 3T}, {\tt deliver}, m, A
\ra$, such that $A$ contains valid signatures for $m$ from $2t+1$
members in $\wt(m)$, and such that $delivery_i[{\sl sender}(m)] = {\sl
seq}(m) - 1$, $p_i$ performs {\sl WAN-deliver}$(m)$ and sets
$delivery_i[{\sl sender}(m)]$ to ${\sl seq}(m)$.  If a timeout period
has passed and $p_j$ is not known to have delivered $m$, $p_i$ sends
$\la {\tt 3T}, {\tt deliver}, m, A \ra$ to $p_j$.

\end{enumerate}
\rule{\columnwidth}{.5mm}
\caption{The {\em 3T} protocol}
\label{fig:3t}
\end{figure}

For each message $m$ the {\em 3T} protocol uses a witness set of size
$2t+1$ out of a potential witness set, $\wt(m)$, of $3t+1$ processes.
The choice of the $2t+1$ threshold is significant, since it guarantees
that a majority of the correct members of $\wt(m)$ acknowledge the
message $m$ and thus no two conflicting messages can receive the
required threshold.  As less than a third of $\wt(m)$ could be
faulty, {\em 3T} ensures Integrity, Reliability, Self-delivery and
Agreement as in the \echo protocol above.  This protocol is used for
failure-recovery in the \act{t} protocol below, in order to guarantee
Self-delivery.

The overhead
incurred (in faultless runs, and not measuring the Stability
Mechanism) is $2t + 1$ signature generations and
message exchanges per delivery.

\section{The \act{t} Protocol}
\label{active}

In this section, we relax the requirements and provide a protocol that
guarantees (only) Probabilistic Agreement.  Thus, we allow the
possibility that a small fraction of the delivered messages
may conflict.

The idea of the \act{t} protocol is as follows: We make use of a
uniformly distributed function $R$ from input pairs of the form $\la
{\sl sender(m)}, {\sl seq(m)} \ra$ designating witness sets $\wact(m)$
of $\kappa$ processes in $P$. The function $R$ is determined at set-up
time, e.g., by seeding it with some random value that processes choose
collectively. By our assumptions, this means that $R$ is unknown to
the adversary in advance, and so the choice of which processes are
faulty is made without knowledge of $R$.  The size of $\wact$,
$\kappa$, is set so that only an exponentially small fraction of the
messages can have a witness set that contains only faulty processes
(who may be collaborating with the sender).  If only $t \le
\onethird{n}$ members are faulty, then by the uniform distribution of $R$,
the expected fraction of messages with a `faulty' witness set is
$\left(\frac{t}{n}\right)^\kappa \le \left(\frac{1}{3}\right)^\kappa$.
Relatively small values of $\kappa$ are sufficient for this to be
negligible.  (As with the size of cryptographic keys, $\kappa$ is
effectively a constant.)  This idea of forming distributed trust in a
cooperation-resilient way borrows from the time-stamping mechanism of
Haber et al.~\cite{HSt91}.  Moreover, we stipulate that correct
processes multicast messages in sequence order, and enforce this
ordering on message delivery. This prevents a malicious sender from
scanning off-line the domain of $\la {\sl sender}, {\sl seq} \ra$
pairs for ones that have faulty witness sets and sending only those
messages.

If $R$ is invertible (an input pair can be easily computed from a
desired output, i.e., a specific, presumably faulty, witness set) then
after $R$ is set, the adversary can compute which are the few
messages it will be able to corrupt. The security of the \act{t}
protocol can be enhanced by the use of a public random oracle $R$,
that maps $\la {\sl sender(m)}, {\sl seq(m)} \ra$ onto $2^P$, such
that the output cannot be distinguished from a random stream. It is
assumed that the oracle can be accessed by all processes and responds
to all queries with one (randomly chosen) mapping. By its randomness
it is implied that the adversary cannot find inputs that map to faulty
process sets. In practice, one adopts the {\em random oracle
methodology}~\cite{FS86,BR93} to approximate $R$, e.g., use a hash
function (such as MD5~\cite{MD5}) in place of $R$, seeded with some
input which is determined at set up time, e.g., by letting the
processes collectively choose a random seed. Note again that by our
assumptions, this implies that the adversary selects which processes
are faulty without knowledge of $R$. Although this is widely
done, we caution the reader that approximating $R$ by a hash function
has no proven security guarantees, and is only heuristically
practically secure~\cite{CGH98}.

\begin{figure}[htb]
\centerline{\input{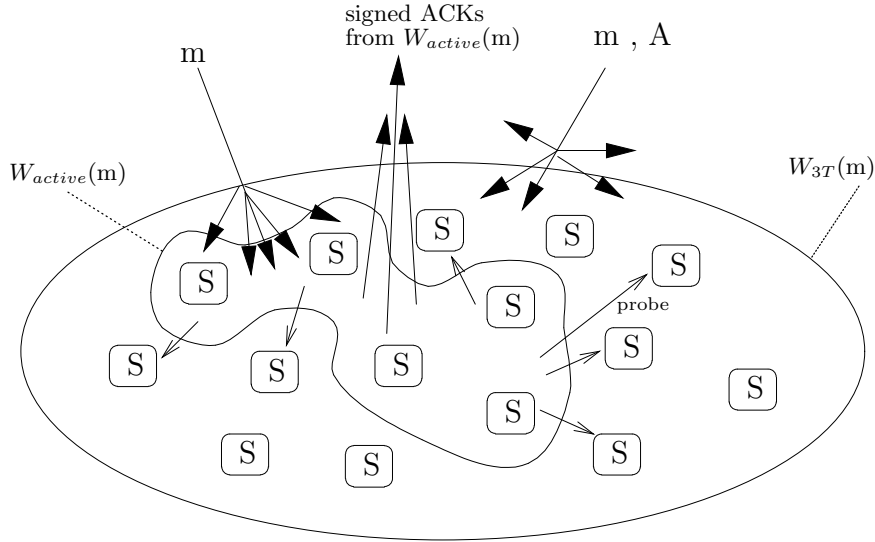}}
\caption{The \act{t} protocol -- no-failure regime}\label{fig:ACT}
\end{figure}

The motivation for this protocol is to choose witness sets
significantly smaller than $2t+1$.  As a result, assuring both safety
and availability is a problem: Since $t$ of the witnesses could be
faulty, for availability one might want to wait for only $\kappa-t$
replies, but usually $\kappa-t < 0$.  Therefore, to guarantee
availability of a witness set for every message, \act{t} incorporates
the {\em 3T} protocol as a recovery-regime.  This is done as follows:
For a process to send a message, it attempts to obtain signed
acknowledgments from $\wact(m)$. We name this the {\em no-failure\/}
regime.  After a timeout period, if $p_i$ has not obtained
acknowledgments from all the members of $\wact(m)$, $p_i$ reverts to
the {\em 3T} protocol and re-sends $m$ to $\wt(m)$ to obtain signed
acknowledgments from a subset of $2t+1$ processes. This is called the
{\em recovery\/} regime.

The integration of the two protocols can potentially create an
opportunity for a faulty process $p_i$ to obtain signed
acknowledgments for conflicting messages.  Specifically, process $p_i$
could first obtain acknowledgements from the small number of witnesses
in $\wact(m)$, then select a different set of $2t+1$ processes to act
as witnesses in the recovery regime.  To decrease such a possibility
of delivering conflicting messages in combining the two regimes, we
provide two measures: First, we turn the witnesses of the no-failure
regime into active participants.  The (correct) members of $\wact(m)$
each probe $\wt(m)$ at $\delta$ randomly chosen {\em peer\/} processes
before acknowledging a message.  Since the peers are chosen by correct
processes during protocol execution, (a correct) one is likely to be
among any $2t+1$ processes chosen to act as witnesses in the recovery
regime.  Figure~\ref{fig:ACT} depicts the active regime
of the \act{t} protocol.  

Secondly, we stipulate that any correct process that receives (signed)
conflicting messages immediately alerts the entire system.  To
guarantee that alerting the system will prevent conflicting messages
from being delivered, in the recovery regime we force a delay before
sending an acknowledgement. By our assumption model, such delay
guarantees, with high probability, that any pending alert message will
arrive at all correct processes.  In practice, this delay can be
reasonably small, e.g., by securing certain bandwidth for control
messages and allowing out-of-band delivery of urgent communication.

In order to allow witnesses to probe their peers on behalf of {\sl
sender}$(m)$, we require every process to sign its own
``acknowledgement-seeking'' ({\sl regular}) messages.  The peer
processes record the message and do not reply if it conflicts with a
previous message. Hence,
knowledge of the message $m$ propagates randomly among correct processes,
without incurring additional signature overhead.  In this way, if a
message $m'$ conflicting with $m$ has been sent to a set $S \subset
\wt(m')$ ($=\wt(m)$) then with high probability the {\em peers} chosen
on behalf of $m$ intersect $S$ at a correct process.  More precisely,
the probability that $\delta$ random probes cross a correct member of
a recovery set containing $2t+1$ processes is at least $1 -
\left(\frac{2t}{3t+1}\right)^\delta \geq 1 -
\left(\frac{2}{3}\right)^\delta$. Therefore, the parameter $\delta$ can be
chosen to achieve any desired level of probabilistic guarantee.

The details of the protocol are given in Figure~\ref{fig:av}.

\noindent
\rule{\columnwidth}{.5mm}
\begin{enumerate}

\item
For a process $p_i$ to {\sl WAN-multicast}($m$), (such that {\sl
sender}$(m) = p_i$), and $p_i$ has previously sent messages up to
sequence number ${\sl seq}(m) - 1$, $p_i$ sends
\[\la {\tt AV}, {\tt regular}, p_i, {\sl seq}(m), H(m), sign \ra
\]
to each $p_j \in \wact(m)$, where $sign = \sign{i}{(p_i, {\sl seq}(m),
H(m))}$.  It then waits to obtain the set of $\kappa$ acknowledgments
$A$ = $\{ \sign{j}{\la {\tt AV}, {\tt ack}, p_i, {\sl seq}(m),
H(m), sign \ra} ~|~ p_j \in \wact(m)
\}$.

If a timeout period has passed and $p_i$ does not obtain 
acknowledgments from all processes in $\wact(m)$, 
then $p_i$ sends 
\[ \la {\tt 3T}, {\tt regular}, p_i, {\sl seq}(m), H(m) \ra
\] 
to $\wt(m)$, and waits to obtain $A$ = 
$\{ \sign{j}{\la {\tt 3T}, {\tt ack}, 
p_i, {\sl seq}(m), H(m) \ra} ~|~ p_j \in P'\}$, 
a set of signed acknowledgments from any subset $P'$ of $\wt(m)$
of $2t+1$ distinct processes.

In either case, $p_i$ then sends to $P$
\[ \la {\tt AV}, {\tt deliver}, m, A \ra~.
\]

\item
When $p_i$ receives a message $\la {\tt AV}$, ${\tt
regular}$, $p_j$, $cnt$, $h$, $sign \ra$, where $sign$ is a valid
signature of $p_j$ on $\la p_j, cnt, h \ra$, it performs the active
phase of secure message transmission: If no conflicting message was
previously received, $p_i$ randomly selects $\delta$ target
processes in $\wt(p_j, cnt)$, denoted \target{i}. It sends
\[
\la {\tt AV}, {\tt inform}, p_j, cnt, h, sign \ra
\]
to every $p_k \in \target{i}$ to obtain a message
$ \la {\tt AV}, {\tt verify}, p_j, cnt, h \ra$
from $p_k$.  
Upon receiving all $\delta$ verifications, 
it then sends to $p_j$ a signed acknowledgment
\[ \sign{i}{\la {\tt AV}, {\tt ack}, p_j, cnt, h, sign \ra}~.
\] 
Note that $p_i$ does not send back to $p_j$ any information about
\target{i}. 

\item
When $p_i$ receives a message $\la {\tt AV}, {\tt inform}, p_j, cnt,
h, sign \ra$ from $p_k$, where $sign$ is a valid signature of $p_j$ on
$\la p_j, cnt, h \ra$, such that no conflicting message was previously
received, it sends to $p_k$
\[ \la {\tt AV}, {\tt verify}, p_j, cnt, h \ra~.
\]

\item 
When $p_i$ receives a message $\la {\tt 3T}, {\tt regular},
p_j, cnt, h \ra$ from $p_j$, such that no conflicting message was
previously received, it delays for a pre-determined timeout period and then
sends to $p_j$ a signed acknowledgment \[ \sign{i}{\la {\tt 3T}, {\tt ack}, p_j, cnt, h \ra}~.
\]

\item
When $p_i$ receives $\la {\tt AV}, {\tt deliver}, m, A \ra$, such that
$A$ contains a valid set of {\tt AV}-acknowledgments from every member
in $\wact(m)$, or a valid set of $2t+1$ {\tt 3T}-acknowledgments from
$\wt(m)$, and such that $delivery_i[{\sl sender}(m)] = {\sl seq}(m) -
1$, $p_i$ performs {\sl WAN-deliver}$(m)$ and sets $delivery_i[{\sl
sender}(m)]$ to ${\sl seq}(m)$.  If a timeout period has passed and
$p_j$ is not known to have delivered a message whose sequence number
is ${\sl seq}(m)$ from ${\sl sender}(m)$, $p_i$ sends $\la {\tt AV}$,
${\tt deliver}$, $m$, $A \ra$ to $p_j$.

\end{enumerate}
\begin{figure}[thb]
\rule{\columnwidth}{.5mm}
\caption{The \act{t} protocol}
\label{fig:av}
\end{figure}

Throughout the protocol, if $p_i$ receives conflicting messages $m$
and $m'$ properly signed by sender $p_j$, $p_i$ immediately sends all processes
alerting message containing $m$ and $m'$, using the fastest
communication channels available to it. The alert message identifies
without doubt a failure in $p_j$ due to the signatures on $m,
m'$. Once $p_j$ is known to have failed, all correct processes
avoid message exchange with $p_j$.  Typically, a malicious
sender may be deterred from sending conflicting messages as their
presence would unquestionably implicate it.

\subsection*{Analysis}
   The \act{t} protocol aims to maximize performance in faultless
cases by minimizing the number of signed messages and the number of
overall message exchanges.  Stress is placed on minimizing digital
signatures (to effectively a constant number per message), since the cost of
producing digital signatures in software is at least one order of
magnitude higher than message-sending, for typical message sizes.

   The overhead in forming agreement on message contents in runs
without failures or pre-mature timeouts is $\kappa$ signature
generations and $\kappa$ message exchanges for collecting $\wact$
acknowledgments and $\delta \times \kappa$ authenticated
message exchanges with {\em peers}. We note that all of the overhead
messages are small (containing fixed size hashes, signatures, and the
like), signatures may be computed concurrently at all of the
witnesses, and likewise, all pairs of message exchanges with peers may
be done concurrently.

   The overhead in case of failures can reach, in the worst case
scenario, $\kappa+3t+1$ signatures and message exchanges with
witnesses of both the no-failure regime and the recovery regime, and
additionally, $\delta \times \kappa$ authenticated message exchanges
between witnesses and their peers. In addition, the recovery regime
incurs a delay on acknowledgement sending to allow for possibly
pending alert messages to reach their destination.

   The level of guarantee achieved by \act{t} depends on the
parameters $\kappa$ and $\delta$. If the system contains as many as
\onethird{n} faulty processes that know about each other (the worst
case scenario), then one out of $3^\kappa$ messages, on average, will have
a completely faulty $\wact$ set.  Since $\wact(m)$ is a function of
{\sl sender}$(m)$ and {\sl seq}$(m)$, whenever $\wact(m)$ has a
completely faulty witness set, {\sl sender}$(m)$ has the opportunity
to collude with the faulty witnesses, and convince correct processes
to {\sl WAN-deliver} conflicting messages.  Moreover, since the
$\wact$ function is known to all participants, once the faulty
processes and the $\wact$ function are determined, the adversary can
predict the sequence number and sender of messages for which it can so
collude and cause conflicting {\sl WAN-deliver} events.  Nonetheless,
this is the case for only an exponentially-small fraction of the
messages that are sent.  By proper choice of the parameter $\kappa$, and
given that messages are multicast in sequence order,
then the likelihood of such a message occurring in the lifetime of the
system can be made appropriately small.

There is also a chance of obtaining acknowledgments signed by correct
members for conflicting messages by having non-intersecting sets of
correct processes participate in the two protocol regimes, as follows:
A faulty process $p_i$ could generate conflicting messages $m$, $m'$,
sent to $\wact(m)$ and $S$ respectively, where $S \subset \wt(m')$, $|S|
= 2t+1$ and $S \cap \wact(m) = \emptyset$. However, if $\wact(m)$
contains at least one correct member $p_h$, then the probability that
\target{h} does not intersect $S$ at a correct member is at most
$\left(\frac{2}{3}\right)^\delta$, which can be made as small as desired
by choosing $\delta$ appropriately.

For example, in a network of $100$ processes, and assuming the number
of faulty processes $t \le 10$, choosing $\kappa = 3$, $\delta = 5$ will
guarantee that conflicting messages are detected with probability at
least $0.95$, whereas in a network of $1000$ processes with $t \le
100$, we can achieve $0.998$ guarantee level with $\kappa = 4$, $\delta = 10$.

\subsection*{Proof of Correctness}

We now proceed to prove Integrity, Self-delivery, Reliability and
Probabilistic Agreement of the \act{t} protocol. We note that due to
the possibility of a completely faulty \wact\/ set, we cannot always
leverage the correctness of the \act{t} protocol from that of the \echo
protocol, as we did in the {\em 3T\/} protocol.

We begin with a statement of several useful
properties of the protocol from which its security leverages.  We note
that in any run of the \act{t} protocol, no correct process multicasts
conflicting messages and no correct process signs conflicting
acknowledgements ({\em 3T\/} or {\em AV}). In addition, we have the
following properties:

\begin{lemma}
\label{lemma2}
In any run of the \act{t} protocol, the following hold:
\begin{enumerate}
\item
\label{lemma2_1}
	If process $p$ is correct in a run, then:
	\begin{enumerate}
	\item
	a correct process signs a {\em 3T\/} acknowledgement 
	for a message $m$ with {\sl sender}(m) = $p$
	only if $p$ multicasts $m$, and
	\item
	a valid signed {\em AV\/} acknowledgement 
	for a message $m$ with {\sl sender}(m) = $p$ can be formed only if
	$p$ multicasts $m$.
	\end{enumerate}

\item
\label{lemma2_2}
If process $p$ is correct in a run, and the run contains a set of valid
{\em 3T\/} or {\em AV\/} acknowledgements for $m$ with {\sl
sender}($m$) = $p$, then $p$ {\sl WAN-multicast} message $m$ in the
run.
\end{enumerate}

\end{lemma}

\begin{proof}
A correct process $q$ signs a {\em 3T\/} acknowledgement for $m$ only when
it receives $m$ over an authenticated channel from {\sl
sender}($m$). Likewise, a correct process signs an {\em AV\/}
acknowledgement only when the {\em AV\/} message contains a valid
signature of {\sl sender}($m$).  The first property then follows from
the fact that a signed acknowledgement of the form $\sign{q}{\la {\tt 3T}, {\tt
ack}, p, cnt, H(m) \ra}$ contains the identity of the sender $p$ = {\sl
sender}($m$), and a signed acknowledgement $\sign{q}{\la {\tt AV}, {\tt
ack}, p, cnt, H(m), sign \ra}$ contains both the sender's identity and
its signature.

For the second property, note that if the run contains a valid set of
{\em 3T\/} acknowledgements for $m$, at least one of which must be
from a correct process, then by item (\ref{lemma2_1}) of the lemma $m$
was multicast by $p$. Likewise, if a valid set of {\em AV\/}
acknowledgements are formed for $m$, then again by item (\ref{lemma2_1}) of
the lemma $m$ was multicast by $p$.
\end{proof}

\begin{teo}(Integrity)
Let $p_i$ be a correct process participating in the \act{t}
protocol. Then $p_i$ executes {\sl WAN-deliver}(m) at most once, and
if {\sl sender}$(m)$ is correct, then only if {\sl sender}$(m)$
executed {\sl WAN-multicast}$(m)$.
\end{teo}

\begin{proof}
That $p_i$ suppresses duplicate deliveries is immediate from the
protocol. It is left to show that if {\sl sender}$(m)$ is correct, it
must have sent $m$.  For $p_i$ to deliver $m$, $p_i$ must have
obtained a valid set of either {\em AV\/} acknowledgments or of {\em
3T\/} acknowledgements for $m$. By Lemma~\ref{lemma2}(\ref{lemma2_2}),
$m$ was {\sl WAN-multicast}$(m)$ by a correct {\sl sender}($m$).
\end{proof}

\begin{teo}(Self-delivery)
Let $p_i$ be a correct process participating in the \act{t}
protocol. If $p_i$ executes {\sl WAN-multicast}(m) then $p_i$ executes
{\sl WAN-deliver}(m).
\end{teo}

\begin{proof}
The theorem easily follows from the Self-delivery property of the
{\em 3T\/} protocol, which is employed within some 
timeout from  {\em WAN-multicast}($m$) unless a valid set of {\em
AV\/} acknowledgements for $m$ is received first, enabling its
delivery by $p_i$.
\end{proof}

\begin{teo}(Reliability)
Let $p_i$ and $p_j$ be two correct processes participating in the \act{t}
protocol. If $p_i$ performs {\sl WAN-deliver}($m$), such that ${\sl
seq}(m) = seq$ and ${\sl sender}(m) = p_k$,
then $p_j$ performs {\sl WAN-deliver}($m'$) such that ${\sl
seq}(m') = seq$ and ${\sl sender}(m') = p_k$.
\end{teo}

\begin{proof}
For $p_i$ to deliver $m$, $p_i$ must have obtained a valid set $A$ of
either {\em AV\/} acknowledgments or of {\em 3T\/} acknowledgements
for $m$.  If $p_i$ learns that $p_j$ delivered $m'$ satisfying ${\sl
seq}(m') = seq$ and ${\sl sender}(m') = p_k$, then by SM\_Integrity we
are done.  Alternatively, if $p_i$ does not learn that $p_j$ delivered
such $m'$, then after a timeout period $p_i$ sends $\la {\tt AV}, {\tt
deliver}, m, A \ra$ to $p_j$.  If $p_j$ has not delivered any
conflicting message, then upon receipt of $\la {\tt AV}, {\tt
deliver}, m, A \ra$ from $p_i$, process $p_j$ performs {\sl
WAN-deliver}($m$). Otherwise, $p_j$ delivers some message $m'$
satisfying ${\sl seq}(m') = seq$ and ${\sl sender}(m') = p_k$.  In
either case, we are done.  (We note that, if {\sl sender}($m$) is
correct, then by Integrity $m = m'$.)
\end{proof}

Note that unlike the \echo and {\em 3T\/} protocols, in the \act{t}
protocol two correct processes are only guaranteed to deliver
the same sequenced message, and not necessarily the same message.
This follows because the protocol only satisfies the Probabilistic
Agreement property, which allows, with some small probability, the
delivery of conflicting messages by different processes.

We now prove that \act{t} maintains Probabilistic Agreement:

\begin{teo}(Probabilistic Agreement)
Let $p_i$ and $p_j$ be two correct processes participating in the
\act{t} protocol. Let $p_i$ deliver a message $m$, $p_j$ deliver $m'$,
such that ${\sl sender}(m) = {\sl sender}(m')$ and ${\sl seq}(m) =
{\sl seq}(m')$. Then the probability that $p_i$ and $p_j$ delivered
conflicting messages, {\em i.e.}, $m \neq m'$, is at most
$\left(\frac{2t}{3t+1}\right)^{\delta} \leq
\left(\frac{2}{3}\right)^{\delta}$\footnote{Here, we take the
probability that an alerting message reaches correct processes in time
to be exactly $1$. By assumption, this probability approximates $1$ as
closely as desired by appropriate tuning of delays.}.
\end{teo}

\begin{proof}
As argued in Theorem~\ref{echoagr} above, for $p_i$ and $p_j$ to
deliver $m$ and $m'$, respectively, they must have each delivered
corresponding sets of valid acknowledgments $A$ and $A'$. Denote by
$\witness{m}$ the set of processes represented in $A$, and likewise
$\witness{m'}$. If $\witness{m}$ and $\witness{m'}$ intersect in an
correct process, then we argue that $m = m'$ as in
Theorem~\ref{echoagr}.  It remains to compute the probability that
conflicting message delivery is enabled in the case that $\witness{m}$
does not intersect $\witness{m'}$ at any correct member.

\begin{description}

\item[Case 1:] $\witness{m} = \witness{m'} = \wact(m)$. Thus,
$\wact(m)$ contains faulty members only. By assumption, the adversary
chooses which processes are faulty without knowledge of $R$, and hence
for any $m$, $\wact(m) = R({\sl sender}(m), {\sl seq}(m))$ randomizes
the choice of processes as a function of $\la {\sl sender(m)}, {\sl
seq(m)} \ra$ independently of failures.  Hence, the probability
$P_\kappa$ for this event is at most $P_\kappa \leq
\left(\frac{t}{n}\right)^\kappa \leq \left(\frac{1}{3}\right)^\kappa$.

\item[Case 2:] $\witness{m} \neq \wact(m)$, $\witness{m'} \neq \wact(m')$. 
Note that $\wt(m)$ $=$ $\wt(m')$, and 
in this case, $\witness{m}, \witness{m'} \subset \wt(m)$ must intersect in a
correct process, leading to a contradiction.

\item[Case 3:] (W.l.o.g.) $\witness{m}
= \wact(m)$, $\witness{m'} \subset \wt(m')$, $|\witness{m'}| =
2t+1$. To distinguish from case 1, assume that $\witness{m}$ contains at
least one correct member $p_h$. Note that the correct member $p_h$
chooses peers randomly, and does not disclose the composition of
\target{h} to {\sl sender}$(m')$. Moreover, by assumption there is a
positive probability for each message from {\sl
sender}$(m')$ to $\wt(m')$ to reach its destination (independent of
the choice of \target{h}).  Thus, the choice of \target{h} is
independent from the choice of any process in
$\witness{m'}$. Therefore, the probability that $p_h$ does not reach
any correct member in $\witness{m'}$ in $\delta$ probes is at most
$\left(\frac{2t}{3t+1}\right)^{\delta} \leq
\left(\frac{2}{3}\right)^{\delta}$.

\end{description}

Thus, the overall probability for conflicting message to be
deliverable is bounded by $\left(\frac{1}{3}\right)^{\kappa} + (1-
\left(\frac{1}{3}\right)^{\kappa})\left(\frac{2}{3}\right)^{\delta}$.
\end{proof}

Obviously, the probability above can be made as small as desired by
appropriate choice of $\kappa, \delta$, for appropriate system sizes
(i.e., such that $n - t \geq \kappa \delta$).

\subsection*{Optimizations}

    It is possible to improve the fault tolerance of the \act{t}
protocol family by allowing any subset of $\kappa-C$ witnesses out of the
designated $\wact$ set, where $C$ is some constant, to validate a
message. Unfortunately, while this improves resilience to benign
failures, it increases the probability of a faulty witness
set. Nevertheless, suppose that $t =
\onethird{n}$. The probability $P_{\kappa,C}$ of a faulty set of $\kappa-C$ out
of $\kappa$ randomly chosen processes is bounded by:

\[
P_{\kappa,C} \approx \sum_{j=0}^{C} 
\frac{ {\frac{n}{3} \choose \kappa-j} {\frac{2n}{3} \choose j} }
	{{n \choose \kappa}} \leq
	\left(\frac{\kappa n}{C(n- \kappa)}\right)^C 
		\left(\frac{1}{3}\right) ^ {\kappa-C} 
\]

This probability tends to zero if we choose $C \ll \kappa$. Therefore, this
allows us to increase the fault tolerance while preserving safety to any
desirable degree.

Similar improvement can be made by accommodating failures in the peer
sets designated by processes in the active probing phase. The details
of the error probabilities induced by such optimizations can be easily
worked out, similarly to the error probability above.

\section{Load}
\label{load}

Our protocols were designed to bring down the cost of forming
agreement on message delivery. This was done by reducing the size of
witness sets used in our protocols.  A related measure of protocol
efficiency is the {\em load\/} it incurs over participating processes,
where by load we mean the expected maximum number of times any server
is accessed per message. To compute load, we need to grow a set $M$ of
randomly selected messages to infinity, and examine the number of
accesses at the busiest server divided by $|M|$.  (This definition is
motivated by Naor and Wool~\cite{NW94}, adapting their definition of
load to our case where distinct messages have different witness
ranges.)  We remark that this definition does not distinguish between
the accesses requiring a server to sign messages and ones requiring it
only to respond.

We first look at the expected access probability of the busiest server
in runs without failures or premature timeouts.  For the $3T$
protocol, the witness sets of messages are subsets of $2t+1$ processes
(each chosen out of a designated range $\wt$ of $3t+1$). If the
$\wt(m)$ function randomizes the choice of processes and likewise,
within every witness range $2t+1$ processes are selected randomly,
then as the number of (randomly selected) messages grows, the
failure-free load on the busiest server tends to $(2t+1)/n$.

In the \act{t} protocol with parameters $\kappa$ and $\delta$, a
message $m$'s delivery involves accessing (in runs without failures or
pre-mature timeouts) a set $\wact(m)$ of $\kappa$ witnesses and a
choice of $\kappa \times \delta$ peer processes.  The choice of
$\wact(m)$ and peers is randomized, giving uniform probabilities for
each process to be accessed when taken at the limit, as the set of
messages goes to infinity. Therefore, the failure-free load of the
\act{t} protocol is $\kappa(\delta + 1) / n$.

If failures occur in the $3T$ protocol, then the load incurred is
bounded by $(3t+1)/n$. This might be acceptable if $t \ll n$.  In the
\act{t} protocol, failures may prevent access to $\wact(m)$ or to
some peers and require accessing some subset of $\wt(m)$ processes for
recovery. 
The load of \act{t} in case of failures is bounded by
$(\kappa(\delta+1) + (3t+1))/ n$.

\section{Conclusions}
\label{conclusion}
Experience in constructing robust distributed
systems~\cite{cacm-gcs,FY91,MR96,Rei95,RFLW96} shows that (secure)
reliable multicast is an important tool for distributed
applications. Implementing reliable multicast in an insecure
environment with arbitrary failures incurs inevitable overhead
required for maintaining consistency. However, a price that may be
acceptable in a small network becomes intolerable for a very large
system.

In this paper, we have shown two approaches in which the requirements
on the system may be weakened in order to allow for more efficient
implementations of reliable multicast: The first is suitable for
environments in which failures are rare, and where therefore, it is
reasonable to assume a low threshold on the number of failures. The
second relaxes the consistency requirement to allow an exponentially small
fraction of the messages to be delivered inconsistently. This approach
is practical when reversing the effects of (a small number of)
bad message deliveries is possible. In both cases,
we have devised protocols that meet the requirements and incur costs
that do not grow with the system size, in normal faultless
scenarios.

\section*{Acknowledgement}

We thank Ran Canetti for discussions on the random oracle methodology
and its limitations. 

\newpage

\end{document}